\documentclass[superscriptaddress,amsmath,amssymb,aps,prx,notitlepage,reprint]{revtex4-1}
\usepackage{graphicx}% Include figure files
\usepackage{dcolumn}% Align table columns on decimal point
\usepackage{bm}% bold math
\usepackage{braket}
\usepackage{amsthm}
\usepackage{epstopdf}
\usepackage{amssymb}
\usepackage{amsmath}
\usepackage{bbold}
\usepackage{hyperref}% add hypertext capabilities
\usepackage{comment}
\usepackage{xcolor}
\theoremstyle{definition}

\begin{document}
\preprint{APS/123-QED}
\title{Optimal distributed quantum sensing using Gaussian states}
\author{Changhun Oh}%
%\email{changhun0218@gmail.com}
\affiliation{Department of Physics and Astronomy, Seoul National University, Seoul 08826, Korea}

\author{Changhyoup Lee}
%\email{changhyoup.lee@gmail.com}
\affiliation{Institute of Theoretical Solid State Physics, Karlsruhe Institute of Technology, Karlsruhe 76131, Germany}

\author{Seok Hyung Lie}%
\affiliation{Department of Physics and Astronomy, Seoul National University, Seoul 08826, Korea}

\author{Hyunseok Jeong}%
%\email{h.jeong37@gmail.com}
\affiliation{Department of Physics and Astronomy, Seoul National University, Seoul 08826, Korea}
\date{\today}% It is always \today, today,

% but any date may be explicitly specified
\begin{abstract}
We find and investigate the optimal scheme of distributed quantum sensing using Gaussian states for estimation of the average of independent phase shifts. 
We show that the ultimate sensitivity is achievable by using an entangled symmetric Gaussian state, which can be generated using a single-mode squeezed vacuum state, a beam-splitter network, and homodyne detection on each output mode in the absence of photon loss. Interestingly, the maximal entanglement of a symmetric Gaussian state is not optimal although the presence of entanglement is advantageous as compared to the case using a product symmetric Gaussian state. It is also demonstrated that when loss occurs, homodyne detection and other types of Gaussian measurements compete for better sensitivity, depending on the amount of loss and properties of a probe state. None of them provide the ultimate sensitivity, indicating that non-Gaussian measurements are required for optimality in lossy cases. Our general results obtained through a full-analytical investigation will offer important perspectives to the future theoretical and experimental study for distributed Gaussian quantum sensing.
\end{abstract}

\maketitle

\section{Introduction}
Quantum resources are known to be useful for further enhancing the precision and the sensitivity of estimation of various physical quantities beyond the standard quantum limit~\cite{giovannetti2004, giovannetti2011, demko2015, pirandola2018, braun2018}. A number of studies on single-parameter estimation have been performed over the last few decades~\cite{Degen2017}, but much attention has begun to be paid to estimation of multiparameters in recent years~\cite{Szczykulska2016}. Quantum-enhanced sensitivity in simultaneous estimation of multiple phases has been investigated to explain the role of quantum entanglement and identify optimal and realistic setups saturating the ultimate theoretical sensitivity~\cite{humphreys2013, Liberman2015, baumgratz2016, Knott2016, pezze2017}. The advantage of exploiting quantum entanglement becomes more significant when sensing takes place in different locations and the parameter of interest is a global feature of the network, e.g., the average of distributed independent phases~\cite{proctor2018, ge2018, guo2019, gessner2018, gatto2019, gessner2019}. Such distributed sensing is related to applications such as global clock synchronization~\cite{komar2014} and phase imaging~\cite{humphreys2013}. These inspire the use of more practical quantum resources that are feasible in a well-controlled manner with current technology, e.g., Gaussian systems~\cite{Gagatsos2016}. Very recently, the sensitivities of distributed quantum sensing with Gaussian states were studied under specific conditions~\cite{guo2019, gatto2019}. The ultimate sensitivity and feasible optimal schemes, however, are not yet found and studied in the class of Gaussian metrology~\cite{ferraro2005, weedbrook2012, adesso2014}.

\begin{figure}[t]
\centering
\includegraphics[width=0.4\textwidth]{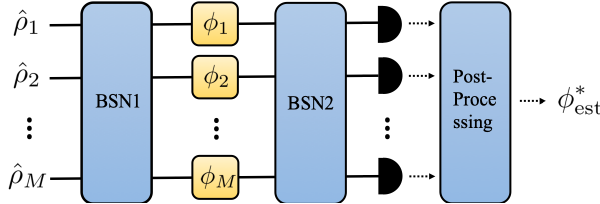}
\caption{Schematic of distributed sensing under investigation.
A multimode probe state $\hat{\rho}_\text{probe}$ generated from the first beam splitter network (BSN) for a given product state input $\otimes_{i=1}^{M}\hat{\rho}_{i}$ undergoes the individual phase shifts on each mode. The parameter-imprinted state $\hat{\rho}_{\bm{\phi}}$ is fed into the second BSN, followed by measurement. The measurement outcomes are used in post-processing to estimate the parameter~$\phi^*=\sum_{i=1}^{M}w_i \phi_i$ with the weight vector~$\bm{w}$.
}
\label{setting}
\end{figure}

%In this Letter, we investigate the achievable sensitivity of Gaussian states \cite{ferraro2005, wang2007, weedbrook2012, adesso2014, serafini2017} in the estimation of the average phases in different sites, as described in Fig.~\ref{setting}.
In this paper, we investigate the ultimate sensitivity for the average phase estimation in distributed quantum sensing with Gaussian states, where the phases are encoded onto a multimode Gaussian probe state, as described in Fig.~\ref{setting}. We find an optimal probe state and measurement setup that achieve the ultimate sensitivity, which are shown to be experimentally feasible with current technology. Interestingly, we demonstrate that the optimal symmetric Gaussian probe state is not a maximally entangled state. For practical relevance, we further analyze the effect of loss, the entanglement-enhanced gain, and other Gaussian measurements in various conditions.

%\paragraph{Gaussian states.---}
We begin with a brief introduction to the formalism describing Gaussian states and multiparameter estimation. Gaussian states are defined as states whose Wigner functions are Gaussian distributions, and thus characterized by the first moment vector~$d_i=\text{Tr}[\hat{\rho}\hat{Q}_i]$ and the covariance matrix~$\Gamma_{ij}=\text{Tr}[\hat{\rho}\{\hat{Q}_i-d_i,\hat{Q}_j-d_j\}/2]$, where~$\{\hat{A},\hat{B}\}\equiv\hat{A}\hat{B}+\hat{B}\hat{A}$. Here, a quadrature operator vector of a~$M$-mode continuous variable quantum system is defined as~$\hat{\bm{Q}}=(\hat{x}_1, \hat{p}_1,...,\hat{x}_M,\hat{p}_M)^\text{T}$, satisfying the canonical commutation relation, $[\hat{Q}_j,\hat{Q}_k]=i(\bm{\Omega}_{2M})_{jk}$, where $\bm{\Omega}_{2M}=
\scriptsize{\begin{pmatrix}
0 & 1 \\
-1 & 0
\end{pmatrix}}\otimes \mathbb{1}_M$ and
%\begin{align*}
%[\hat{Q}_j,\hat{Q}_k]=i(\bm{\Omega}_{2M})_{jk}, ~~~\bm{\Omega}_{2M}=
%\begin{pmatrix}
%0 & 1 \\
%-1 & 0
%\end{pmatrix}\otimes \mathbb{1}_M,
%\end{align*}
$\mathbb{1}_{M}$ is the $M \times M$ identity matrix.

\section{Distributed sensing}
Consider estimation of $M$-parameter~$\bm{\phi}=(\phi_1,\phi_2,...,\phi_M)^\text{T}$ based on measurement outcomes~$\bm{x}$, obtained with a conditional probability~$p(\bm{x}\vert\bm{\phi})$.
The multiparameter Cram\'{e}r-Rao inequality states that the~$M\times M$ estimation error matrix~$\Sigma_{ij}=\langle (\hat{\phi}_i-\phi_i)(\hat{\phi}_j-\phi_j) \rangle$ of any unbiased estimator~$\hat{\phi}_i$ is bounded by the Fisher information matrix (FIM),~$\bm{F}(\bm{\phi})$, i.e.,~$\bm{\Sigma} \geq \bm{F}^{-1},$ where~
$\bm{F}_{ij}(\bm{\phi})=\sum_{\bm{x}} \frac{1}{p(\bm{x}\vert\bm{\phi})}\frac{\partial p(\bm{x}\vert \bm{\phi})}{\partial\phi_i}\frac{\partial p(\bm{x}\vert\bm{\phi})}{\partial\phi_j}$
~\cite{helstrom1976}.
The conditional probability~$p(\bm{x}\vert\bm{\phi})=\text{Tr}[\hat{\rho}_{\bm{\phi}} \hat{\Pi}_{\bm{x}}]$ is given by a positive operator-valued measure~$\hat{\Pi}_{\bm{x}}$ for a given parameter-encoded state~$\hat{\rho}_{\bm{\phi}}$.
The quantum Cram\'{e}r-Rao inequality sets a lower bound for the error of an unbiased estimator, i.e., $\bm{\Sigma} \geq \bm{F}^{-1} \geq \bm{H}^{-1}$,
where~$H_{ij}=\text{Tr}[\hat{\rho}_{\bm{\phi}}\{\hat{L}_i,\hat{L}_j\}]/2$ is the quantum Fisher information matrix (QFIM), with~$\hat{L}_i$ being a symmetric logarithmic derivative operator associated with~$i$th parameter~$\phi_i$ \cite{braunstein1994}.
When a linear combination of~$\phi_i$'s,~$\phi^*=\bm{w}^\text{T}\bm{\phi}=\sum_{i=1}^M w_i\phi_i$ with the weight vector~$\bm{w}$, is of particular interest, the estimation error is bounded as \cite{paris2009}
\begin{align}\label{err}
\Delta^2\phi^*\equiv \langle (\hat{\phi}^*-\phi^*)^2\rangle\geq \bm{w}^\text{T}\bm{F}^{-1}\bm{w}\geq \bm{w}^\text{T}\bm{H}^{-1}\bm{w}.
\end{align}
Here,~$\bm{F}^{-1}$ and~$\bm{H}^{-1}$ are understood as the inverse on their support if the matrices are singular.
Throughout this paper, we assume the normalization~$\sum_{i=1}^M\vert w_i\vert=1$ for simplicity.

\section{Gaussian distributed sensor}
\subsection{Quantum Fisher information matrix}
Consider a distributed phase sensor in which a product Gaussian input state~$\otimes_{i=1}^{M}\hat{\rho}_i$ is injected into a beam splitter network (BSN), preparing a probe state $\hat{\rho}_\text{probe}$, the multiphase information is encoded onto $\hat{\rho}_\text{probe}$ by a unitary operation~$\hat{U}_{\bm{\phi}}=\exp(-i\sum_{j=1}^M \phi_j \hat{a}_j^\dagger \hat{a}_j)$, and the output state~$\hat{\rho}_{\bm{\phi}}$ is measured after the second BSN, as depicted in~Fig.~\ref{setting}.
Note that configuration of the first BSN enables one to generate any probe Gaussian states \cite{Reck1994, weedbrook2012}. We also implicitly assume a strong reference beam to define the phases, accessible in each mode for measurement~\cite{jarzyna2012}.
Here, we aim to investigate the sensitivity of Gaussian states for estimation of the parameter~$\phi^*$. 
When the probe state $\hat{\rho}_\text{probe}$ after the first BSN is a pure Gaussian state characterized by~$(\bm{\Gamma}, \bm{d})$, the elements of the QFIM are written as \cite{banchi2015, serafini2017, nichols2018, oh2019-2, liu2019, sidhu2019}
\begin{align}\label{qfim}
H_{ij}=&2\text{Tr}[\bm{\Gamma}_\text{probe}^{(i,j)}\bm{\Gamma}_\text{probe}^{(j,i)}]-\delta_{ij}+(\bm{\Omega}_2 \bm{d}_\text{probe}^{(i)})^\text{T}[\bm{\Gamma}_\text{probe}^{-1}]^{(i,j)} \nonumber
\\&\times (\bm{\Omega}_2 \bm{d}_\text{probe}^{(j)}),
\end{align}
where~$\bm{A}^{(i,j)}$ denotes the~$2\times 2$ submatrix in the $i$th row and~$j$th column of the $M\times M$ block matrix~$\bm{A}$, and similar for the vector~$\bm{d}^{(i)}$.
The derivation of the QFIM of Eq.~\eqref{qfim} is provided in Appendix A.
The convexity of QFIM makes it sufficient to consider only pure probe states to find an optimal state maximizing the QFIM~\cite{ge2018}, but one can find the analytical form of the QFIM for general Gaussian states~\cite{banchi2015, serafini2017, nichols2018, oh2019-2, liu2019, sidhu2019}.
The quantum Cram\'{e}r-Rao bound in Eq.~\eqref{err} can be saturated since the generators of parameters commute~\cite{pezze2017}.
%, i.e.,~$[\hat{a}^\dagger_i\hat{a}_i,\hat{a}^\dagger_j\hat{a}_j]=0$ for any~$i,j$ pairs~\cite{pezze2017}.

\subsection{Optimal product Gaussian state}
Let us first consider the case where the probe state is a product state and thus the QFIM is evidently a diagonal block matrix.
Without loss of generality, we assume that the block matrix of the covariance matrix for~$i$th mode is~$\bm{\Gamma}^{(i,i)}=\text{diag}(e^{2r_i},e^{-2r_i})/2$, simplifying the estimation error of~$\phi^*$ to be
%\begin{align*}
$\Delta^2\phi^*\geq \sum_{i=1}^M w_i^2/(\cosh{4r_i}-1+2d_{2i}^2 e^{-2r_i}+2d_{2i-1}^2 e^{2r_i})$.
%\end{align*}

When probing with a product coherent state, the error bound becomes~$\sum_{i=1}^M w_i^2/2 (d_{2i}^2+d_{2i-1}^2)$,
and the best strategy for a given total average photon number~$\bar{N}$ is to distribute the energy $\bar{N}$ over the modes according to the weight~$\vert w_i\vert$, i.e.,~$\bar{N}_i=(d_{2i}^2+d_{2i-1}^2)/2=\bar{N} \vert w_i\vert$.
The estimation error is thus
\begin{align*}
\Delta^2\phi^*\geq \sum_{i=1}^M \frac{w_i^2}{4\bar{N}_i}=\frac{1}{4\bar{N}}\equiv \Delta^2\phi^*_\text{SQL},
\end{align*}
where the lower bound defines the standard quantum limit. When~$w_i=1/M$, i.e., $\phi^*$ is the average phase, $\Delta^2\phi^*_\text{SQL}=1/4M\bar{n}$, where $\bar{n}\equiv\bar{N}/M$ represents an equal average number of photons hitting each phase shifter.

\begin{figure}[b]
\centering
\includegraphics[width=0.4\textwidth]{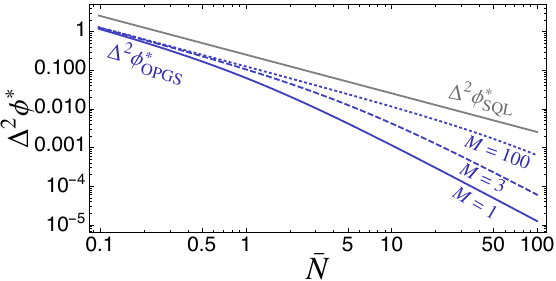}
\caption{Estimation errors when probing the phases with product states. 
The top curve represents the standard quantum limit, $\Delta^2\phi^*_\text{SQL}$, whereas the other curves show the error $\Delta^2\phi^*_\text{OPGS}$ when $M=1, 3, 100$. 
The error $\Delta^2\phi^*_\text{OPGS}$ increases with $M$ for a fixed~$\bar{N}$, but is always below the error $\Delta^2\phi^*_\text{SQL}$, which approaches $1/8\bar{N}$ as $M\rightarrow \infty$.
}
\label{fig:prod}
\end{figure}

Among all product Gaussian states, the best strategy under the energy constraint~$\bar{N}$ 
%$\bar{N}=\sum_{i=1}^M \sinh^2r_i +|\bm{d}|^2$ 
is to prepare the probe state in a product squeezed vacuum state with~$8\bar{N}_i^2(\bar{N}_i+1)/(2\bar{N}_i+1)\propto w_i^2$.
Thus, particularly when~$w_i=1/M$, in which~$\phi^*$ is the average phase, the estimation error becomes
\begin{align}\label{OPGSerror}
\Delta^2\phi^*\geq\frac{M}{8\bar{N}(\bar{N}+M)}=\frac{1}{8M\bar{n}(\bar{n}+1)}\equiv \Delta^2\phi^*_\text{OPGS},
\end{align}
where we have set~$r_i=r$ for all~$i$ and~$\bar{N}=M \sinh^2 r$. Note that the Heisenberg scaling with~$\bar{n}$ or $\bar{N}$ is achieved. We refer to the above product squeezed vacuum state as \textit{the optimal product Gaussian state} (OPGS) throughout this paper.
The error $\Delta^2\phi^*_\text{OPGS}$ grows with the number of modes $M$, over which the probe state is distributed for a given~$\bar{N}$, as shown in~Fig.~\ref{fig:prod}.
When an equal energy can be used in all the modes, i.e., for a fixed~$\bar{n}$, the error $\Delta^2\phi^*_\text{OPGS}$ decreases with~$M$, which is obvious since the total energy being used increases by $M$ times.
It can be easily shown that the estimation error $\Delta^2\phi^*_\text{OPGS}$ can be achieved by performing homodyne detection on each mode without the second BSN \cite{olivares2009}.

\subsection{Optimal entangled Gaussian state}
We now turn to the case when the first BSN is configured to create mode correlation for an injected product input state. In order to find the ultimate sensitivity in distributed sensing using Gaussian states and an optimal probe state, one can further develop the inequality of Eq.~\eqref{err} as 
\begin{align}
\Delta^2 \phi^*&\geq\bm{w}^\text{T}\bm{H}^{-1}\bm{w}\geq\frac{|\bm{w}|^4}{\bm{w}^\text{T}\bm{H}\bm{w}}=\frac{|\bm{w}|^4}{4(\Delta^2 \hat{G}^*)_\psi}\nonumber \\
&\geq \frac{|\bm{w}|^4}{4 \max_\psi (\Delta^2 \hat{G}^*)_\psi}, \nonumber
\end{align}
where $\hat{G}^*=\sum_{i=1}^M w_i \hat{a}_i^\dagger \hat{a}_i$ is the generator of~$\phi^*$~\cite{proctor2018}.
From now on, let us focus on the estimation of the average phase, i.e.,~$w_i=1/M$.
%, for which the generator is given by~$\hat{G}^*=\sum_i \hat{a}_i^\dagger\hat{a}_i$.
Using a series of inequalities, we show that the error for the average phase estimation is given by (see Appendix B for the detail)
\begin{align}
%\label{ult}
\Delta^2\phi^*\geq \frac{1}{8\bar{N}(\bar{N}+1)}=\frac{1}{8M\bar{n}(M\bar{n}+1)}\equiv \Delta^2\phi^*_\text{OEGS},\nonumber
\end{align}
We note that the ultimate error $\Delta^2\phi^*_\text{OEGS}$ scales with~$\bar{N}^{-2}$ or $\bar{n}^{-2}$, and is smaller than the error $\Delta^2\phi^*_\text{OPGS}$.
A similar scaling has been discussed in Ref.~\cite{ge2018}, but with different quantification of the resource.

%\paragraph{Symmetric Gaussian states for optimality.---}
We show that the ultimate error $\Delta^2\phi^*_\text{OEGS}$ can be achieved by using symmetric Gaussian probe states with zero displacement. 
The covariance matrix of pure symmetric Gaussian probe states can be written as 
a $M\times M$ partitioned matrix $\bm{\Gamma}_\text{probe}$ with submatrices $\bm{\Gamma}_\text{probe}^{(i,i)}=\text{diag}(\gamma_1,\gamma_2)$ for all~$i$ and~$\bm{\Gamma}_\text{probe}^{(i,j)}=\text{diag}(\epsilon_1,\epsilon_2)$ for all~$i\neq j$~\cite{adesso2004, serafini2005, serafini2017, adesso2008}.
Since the states are assumed to be pure, the components obey the relations~$(\gamma_1-\epsilon_1)(\gamma_2-\epsilon_2)=1/4$ and~$[\gamma_1+(M-1)\epsilon_1][\gamma_2+(M-1)\epsilon_2]=1/4$ \cite{adesso2004, serafini2005, adesso2008}.
The QFIM for symmetric Gaussian states is evidently a symmetric matrix with $H_{ii}=H_{11}$ for all~$i$ and $H_{ij}=H_{12}$ for all~$i\neq j$. 
Finally, using Eqs.~\eqref{err} and \eqref{qfim}, the estimation error is reduced to
\begin{align}\label{merror}
\Delta^2\phi^*\geq \frac{1}{M[H_{11}+(M-1)H_{12}]},
\end{align}
where~$H_{11}=2(\gamma_1^2+\gamma_2^2)-1$ and~$H_{12}=2(\epsilon_1^2+\epsilon_2^2)$ (see Appendix C for details).
It is clear that the correlation quantified by~$\epsilon_1$ and~$\epsilon_2$,~$H_{12}$, plays an important role, but the sensitivity is eventually determined by an interplay with the term~$H_{11}$ that is not independent of~$H_{12}$ for a given energy. 
After minimizing the lower bound in Eq.~\eqref{merror} under the energy constraint~$\bar{N}=M(\gamma_1+\gamma_2-1)/2$ (see Appendix D for the detail), we recover the ultimate error $\Delta^2\phi^*_\text{OEGS}$ when 
$\gamma_{1,2}=1/2+\epsilon_{1,2}$ and
$\epsilon_{1,2}=[\bar{N}\pm\sqrt{\bar{N}(\bar{N}+1)}]/M$, leading to
$H_{11}=4 \bar{N} (2 \bar{N}+M+1)/M^2$ and~$H_{12}=4\bar{N}(2\bar{N}+1)/M^2$.
Therefore, the ultimate estimation error~$\Delta^2\phi^*_\text{OEGS}$ can be achieved by the optimal symmetric Gaussian state, which we call \textit{the optimal entangled Gaussian state} (OEGS) throughout this paper.
Most importantly, in contrast to the error $\Delta^2\phi^*_\text{OPGS}$, the error $\Delta^2\phi^*_\text{OEGS}$ is independent of the number of modes $M$ for a fixed energy~$\bar{N}$ and scales with~$M^{-2}$ for a fixed~$\bar{n}$, evidently resulting from exploiting entanglement.
Thus, the mode entanglement enables one to prevent the estimation error from growing with $M$.

\begin{figure}[b]
\centering
\includegraphics[width=0.42\textwidth]{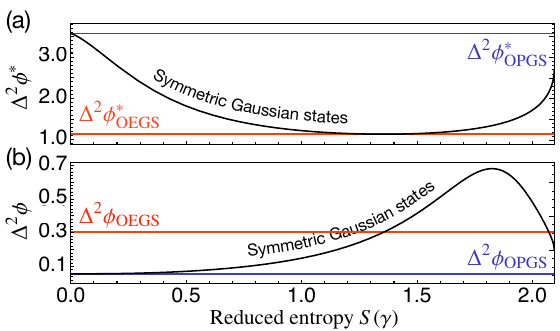}
\caption{Dependence of the reduced entropy $S(\bm{\gamma})$ in the estimation error when probing with symmetric Gaussian states (black curves) in comparison with the OPGS (blue lines) and the OEGS (red lines). 
(a) For the average phase estimation, the OEGS achieving the ultimate sensitivity does not have the maximal reduced entropy. (b) However, the OPGS is optimal for the simultaneous phase estimation.  
}
\label{OEGS_plot}
\end{figure}

\subsection{Role of entanglement}
One might wonder whether the OEGS is the maximally entangled Gaussian state, for which the entropy of the reduced state is maximized.
We now demonstrate that it is not the case.
The entropy of the single-mode reduced state having a diagonal covariance matrix~$\bm{\gamma}$ is given by
$S(\bm{\gamma})=\bar{n}_\text{T} \ln(1+1/\bar{n}_\text{T})+\ln(\bar{n}_\text{T}+1)$ \cite{ferraro2005},
where~$\bar{n}_\text{T}=\sqrt{\gamma_1\gamma_2}-1/2$ is the average thermal photon number of the reduced single-mode state.
The entropy~$S(\bm{\gamma})$ increases with the entanglement of the total system under investigation, where pure symmetric Gaussian states are only considered \cite{adesso2005}.
Interestingly, the OEGS achieving the ultimate sensitivity does not have the maximal entropy, as shown in Fig.~\ref{OEGS_plot}(a). This is surprising and in contrast to other cases, where maximally entangled states have shown to lead to the optimal sensitivity, e.g., the GHZ state of qubits exhibiting the maximal entropy of the reduced state~\cite{proctor2018}.
In our scenario, the state often referred to as the continuous variable GHZ-type state having the maximal reduced entropy~\cite{loock2000, loock2003, yonezawa2004} exhibits worse sensitivity than the OEGS. A similar result has been reported for estimation of unitarily generated parameters in Ref.~\cite{baumgratz2016}.

%\paragraph{Simultaneous estimation error.---}
It is worth comparing with the error of simultaneous phase estimation,~$\Delta^2\phi\equiv\sum_{i=1}^M \Delta^2\phi_i$.
For general symmetric Gaussian states without displacement, the error can be written as $\Delta^2\phi\geq\text{Tr}[\bm{H}^{-1}]=(M-1)/(H_{11}-H_{12})+1/[H_{11}+(M-1)H_{12}]$,
%\begin{align}\label{tot_eq}
%\Delta^2\phi\geq\text{Tr}[\bm{H}^{-1}]&=\frac{M-1}{H_{11}-H_{12}}+\frac{1}{H_{11}+(M-1)H_{12}},
%\end{align}
where the first term will be ignored if~$H_{11}=H_{12}$.
For a product probe state,~$H_{12}$ disappears and thus,
\begin{align} 
%\label{tot_err_prod}
\Delta^2\phi\geq \frac{M^3}{8\bar{N}(\bar{N}+M)}
%=\frac{M}{8\bar{n}(\bar{n}+1)}
\equiv \Delta^2\phi_\text{OPGS},\nonumber
\end{align}
where the bound $\Delta^2\phi^*_\text{OPGS}$ can be achieved by the OPGS.
When using the OEGS, however, the estimation error is given by
\begin{align} 
%\label{tot_err_ent}
\Delta^2\phi&\geq\frac{M[2\bar{N}(M-1)+2M-1]}{8\bar{N}(\bar{N}+1)} 
%\nonumber \\ 
%&=\frac{2M\bar{n}(M-1)+2M-1}{8\bar{n}(M\bar{n}+1)}
\equiv\Delta^2\phi_\text{OEGS}. \nonumber
\end{align}
It is clear that the error $\Delta^2\phi_\text{OEGS}$ is larger than the error $\Delta^2\phi_\text{OPGS}$. More generally, any entangled symmetric Gaussian states exhibit worse sensitivity than the OPGS, as shown in Fig.~\ref{OEGS_plot}(b).

\section{Practical Perspectives}
\subsection{Physical implementation of the optimal scheme}
We have shown above that the ultimate estimation error $\Delta^2\phi^*_\text{OEGS}$ is achieved by the OEGS. Generation of the latter is experimentally feasible with current technology as we provide here. Suppose that a product state of a~$p$-squeezed vacuum and~$(M-1)$ vacua is injected into the first BSN, configured as
$\hat{U}_\text{BSN}=\hat{B}_{M-1,M}(\theta_{M-1})\hat{B}_{M-2,M-1}(\theta_{M-2})\times \cdot\cdot\cdot \times \hat{B}_{1, 2}(\theta_1)$,
where $\hat{B}_{i,j}(\theta_j)=\exp[\theta_j(\hat{a}_i^{\dagger}\hat{a}_{j}-\hat{a}_i\hat{a}_{j}^{\dagger})]$ and $\theta_j=\arccos(M-j+1)^{-1/2}$. 
Consequently, one can show that the output state of the BSN is the OEGS. Notice that different configurations of BSN can be employed to generate the OEGS~\cite{guo2019}.
%The $\hat{U}_\text{BSN}$ with an input of a~$x$-squeezed vacuum and~$(M-1)$~$p$-squeezed vacua with an equal squeezing strength can also generate a GHZ-type entangled state having maximal reduced entropy~\cite{loock2000, loock2003}, but it shows worse sensitivity than the OEGS. 

%\paragraph{Optimal measurement.---}
We demonstrate here that homodyne detection on each mode is sufficient to achieve the ultimate error $\Delta^2\phi^*_\text{OEGS}$ without using the second BSN. The resultant probability distribution of homodyne detection follows a Gaussian distribution with the zero first-moment vector and the~$M\times M$ covariance matrix~$\bm{\Gamma}_\text{HD}$ with diagonal components~$[\bm{\Gamma}_\text{HD}]_{ii}=\gamma_1 \cos^2\varphi_i+\gamma_2 \sin^2\varphi_i$ and off-diagonal components~$[\bm{\Gamma}_\text{HD}]_{ij}=\epsilon_1 \cos\varphi_i\cos\varphi_j+\epsilon_2 \sin\varphi_i\sin\varphi_j$ where~$\varphi_i=\phi_i-\theta_{\text{HD},i}$ with homodyne angles~$\theta_{\text{HD},i}$ on~$i$th mode.
The error is thus given by~$\Delta^2\phi^*_\text{HD}\geq\bm{w}^\text{T}\bm{F}^{-1}\bm{w}$, where~$F_{ij}=\text{Tr}[\bm{\Gamma}_\text{HD}^{-1}(\partial_{\phi_i}\bm{\Gamma}_\text{HD})\bm{\Gamma}_\text{HD}^{-1}(\partial_{\phi_j}\bm{\Gamma}_\text{HD})]/2$.
It can be easily shown that the lower bound is equal to $\Delta^2\phi^*_\text{OEGS}$ when~$\varphi_i=\varphi_\text{opt}\equiv\pi/2-\cot^{-1}[2\sqrt{\bar{N}(\bar{N}+1)}]/2$ for all~$i$.
Such optimal phase setting can be made by adjusting the homodyne angles~$\theta_{\text{HD},i}=\phi_i-\varphi_\text{opt}$.
%Hence, by tuning the homodyne angles on each mode, one can achieve the optimal sensitivity of Eq.~\eqref{ult}.

\begin{figure}[t]
\centering
\includegraphics[width=0.48\textwidth]{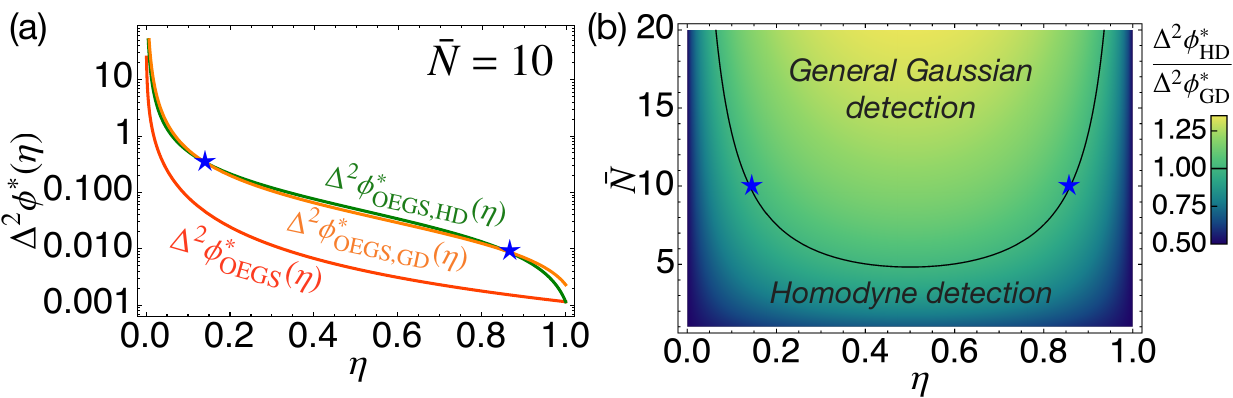}
\caption{(a)~Comparison among the estimation errors $\Delta^2\phi^*_\text{OEGS}(\eta)$ (red curve), $\Delta^2\phi^*_\text{OEGS,HD}(\eta)$ (green curve), and $\Delta^2\phi^*_\text{OEGS,GD}(\eta)$ (orange curve) with loss $\eta$ for $\bar{N}=10$. 
The estimation error of the optimal scheme $\Delta^2\phi^*_\text{OEGS}(\eta)$ achieves an improvement by an order of magnitude compared to homodyne detection and general Gaussian detection when loss is significant.
Note that there exist two crossing points ($\star$) between $\Delta^2\phi^*_\text{OEGS,HD}(\eta)$ and $\Delta^2\phi^*_\text{OEGS,GD}(\eta)$ as $\eta$ increases.
(b)~The ratio of $\Delta^2\phi^*_\text{OEGS,HD}(\eta)$ to $\Delta^2\phi^*_\text{OEGS,GD}(\eta)$ as a function of $\bar{N}$ and $\eta$. The boundary, represented by a solid line, is given by $\bar{N}=(1+\sqrt{2})/2\eta(1-\eta)$, at which homodyne detection and general Gaussian detection after the second BSN ($\hat{U}_\text{BSN}^{-1}$) yields the same sensitivity. General Gaussian detection scheme becomes significant only when $\bar{N}>2(1+\sqrt{2})$, and exhibits the most advantage over the homodyne detection at $\eta=0.5$.}
\label{fig:fig4}
\end{figure}

\subsection{Effects of loss}
From a practical perspective, we analyze the effect of photon loss on the sensitivity. When loss is assumed to occur in each mode with an equal $\eta$, the covariance matrix of the probe state is transformed as~$\bm{\Gamma}_\text{probe}\rightarrow\eta\bm{\Gamma}_\text{probe}+(1-\eta)\mathbb{1}_{2M}/2$, i.e., $\gamma_{1,2}\rightarrow \eta\gamma_{1,2}+(1-\eta)/2$ and~$\epsilon_{1,2}\rightarrow \eta \epsilon_{1,2}$~\cite{ferraro2005, serafini2017}. Consequently, the theoretical optimal error bounds $\Delta^2\phi^*_\text{OPGS}$ and $\Delta^2\phi^*_\text{OEGS}$ become
\begin{align}
\Delta^2\phi^*_{\text{OPGS}}(\eta)&\equiv 1/4\bar{N}\eta(2\bar{N}\eta/M+\eta+1), \nonumber\\
%\label{loss_prod} 
\Delta^2\phi^*_{\text{OEGS}}(\eta)&\equiv 1/4\bar{N}\eta(2\bar{N}\eta+\eta+1), \nonumber
%\label{loss_ent},
\end{align}
respectively. When homodyne detection is performed, the resulting error bounds are respectively given as
\begin{align}
\Delta^2\phi^*_{\text{OPGS,HD}}(\eta)&\equiv [4\bar{N}\eta(1-\eta)+M]/[8\eta^2\bar{N}(\bar{N}+M)],\nonumber\\
%\label{hom_prod} \\
\Delta^2\phi^*_{\text{OEGS,HD}}(\eta)&\equiv [4\bar{N}\eta(1-\eta)+1]/[8\eta^2\bar{N}(\bar{N}+1)], \nonumber
%\label{hom_ent}
\end{align}
for which the homodyne angles have been appropriately chosen. 
One may also seek other type of Gaussian measurement that could outperform the case yielding $\Delta^2\phi^*_{\text{OEGS,HD}}(\eta)$ in the presence of loss.
We exemplify the latter by performing an appropriate general-dyne detection on the first output mode and heterodyne detection on the other output modes of the second BSN that is set to realize $\hat{U}_\text{BSN}^{-1}$. The associated error bound when probing with the OEGS is given as
\begin{align}
\Delta^2\phi^*_\text{OEGS,GD}(\eta) \equiv \frac{2\bar{N} (1-\eta) \eta+1+\sqrt{1-4\bar{N} \eta(\eta -1)}}{8 \eta ^2 \bar{N} (\bar{N}+1)},\nonumber
\end{align}
whose derivation and detailed setup are provided in Appendix E. Figure~\ref{fig:fig4}(a) reveals that the error $\Delta^2\phi^*_\text{OEGS,HD}(\eta)$ is competitive with $\Delta^2\phi^*_\text{OEGS,GD}(\eta)$ depending on $\eta$, and none of them attain the ultimate error $\Delta^2\phi^*_\text{OEGS}(\eta)$ when $\eta<1$. Comparable behaviors between $\Delta^2\phi^*_\text{OEGS,HD}(\eta)$ and $\Delta^2\phi^*_\text{OEGS,GD}(\eta)$ are elaborated in terms of $\bar{N}$ and $\eta$ in Fig.~\ref{fig:fig4}(b), identifying the regions in which one prevails over the other. It shows that homodyne detection is advantageous when $\bar{N}>(1+\sqrt{2})/2\eta(1-\eta)$. 
Interestingly, the error bound $\Delta^2\phi^*_\text{OEGS,HD}(\eta)$ is exactly the same as that of a single-mode phase estimation using a squeezed thermal state~\cite{oh2019-1}.
One could further reduce the error by having displacement as in Ref.~\cite{guo2019}, or seek for non-Gaussian measurements to achieve the ultimate error $\Delta^2\phi^*_\text{OEGS}(\eta)$ in lossy cases~\cite{oh2019-1, oh2019-2}.

\begin{figure}[t]
\centering
\includegraphics[width=0.48\textwidth]{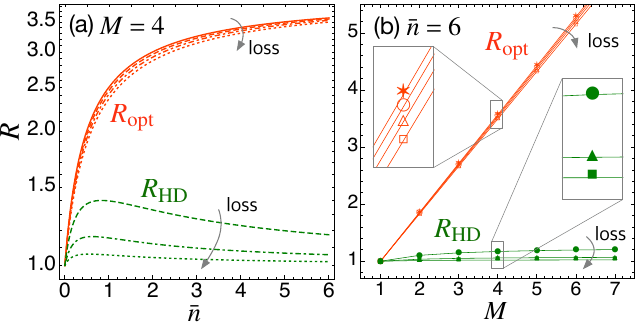}
\caption{The relative error ratios $R_\text{opt}$ and $R_\text{HD}$ for the cases with $\eta=1~($solid$), 0.9~($dashed$), 0.8~($dot-dashed$), 0.7~($dotted$)$
 (a) as a function of $\bar{n}$ when $M=4$, and  (b) as a function of $M$ when $\bar{n}=6$.
The quantum enhancement offered by the optimal scheme is robust against photon loss.
Overall, loss is obviously always detrimental for a given $\bar{n}$ and $M$, i.e., the error ratios decrease with~$\eta$. Lines connecting dots in panel (b) are to guide the eyes.
}
\label{fig:final}
\end{figure}

%\paragraph{Entanglement-enhanced sensitivity.---}
The enhancement of sensitivity by entanglement can be quantified by the relative error ratio $R_\text{opt}=\Delta^2 \phi^*_\text{OPGS}(\eta)/\Delta^2 \phi^*_\text{OEGS}(\eta)$ for the case that an optimal measurement is assumed, and the error ratio $R_\text{HD}=\Delta^2 \phi^*_\text{OPGS,HD}(\eta)/\Delta^2 \phi^*_\text{OEGS,HD}(\eta)$ for the case that homodyne detection is performed. 
%\begin{align}
%R=\frac{(\Delta^2 \phi^*_\text{OEGS})^{-1}}{(\Delta^2 \phi^*_\text{OPGS})^{-1}}&=\frac{2\bar{N}\eta+\eta+1}{2\bar{N}\eta/M+\eta+1}=\frac{2M\bar{n}\eta+\eta+1}{2\bar{n}\eta+\eta+1},
%\end{align}
%which becomes~$M(\bar{N}+1)/(\bar{N}+M)=(M\bar{n}+1)/(\bar{n}+1)$ when photon-loss does not occur, i.e.~$\eta=1$.
Figure~\ref{fig:final}(a) shows that the $R_\text{opt}$ slightly decreases with a moderate loss~$\eta$ and monotonically increases with~$\bar{n}$, while the $R_\text{HD}$ drastically drops with $\eta$ and exhibits the optimum at $\bar{n}=1/2\sqrt{M\eta(1-\eta)}$, where the relative enhancement is maximal, when $\eta<1$.
The behaviors of $R_\text{opt}$ and $R_\text{HD}$ with increasing $M$ are presented in Fig.~\ref{fig:final}(b) for $\bar{n}=6$. 
Remarkably, both $R_\text{opt}$ and $R_\text{HD}$ are always greater than unity in all cases with any $\eta$, stressing the usefulness of entanglement in Gaussian distributed sensing against loss.

%Thus,it is required to use a joint non-Gaussian measurement composed of the projections onto the eigenbasis of the symmetric logarithmic derivative~$\hat{L}_{\phi^*}$ \cite{braunstein1994, oh2019-2}; for example, for~$\phi^*=0$,~$\hat{L}_{\phi^*}\propto\hat{Q}^\text{T}(\Omega_{2M}\Gamma+\Gamma \Omega_{2M})\hat{Q}$.

\section{Discussion}
We have investigated the ultimate sensitivity of the average phase estimation in distributed quantum sensing using Gaussian states.
The ultimate sensitivity has been shown to be achievable by the OEGS possessing partial entanglement between the modes and by performing homodyne detection on each mode in the absence of loss. 
When photon loss occurs, homodyne detection ceases to be optimal, but non-Gaussian measurement would be required for achieving the ultimate sensitivity. 
Alternatively, a slightly better sensitivity can be obtained by conducting other type of Gaussian measurement on the output modes of the second BSN that implements the inverse transformation of the first BSN. Although the sensitivity decreases with loss in all the cases considered in this work, we have revealed that using the OEGS is always advantageous for average phase estimation as compared to the case using unentangled symmetric Gaussian states. 
While we have focused on identification of the ultimate sensitivity and the optimal setup for the average phase estimation in this work, finding those for estimation of other linear combinations of phases would also be an interesting future study. 
Another interesting open question is to explain the enhancement of sensitivity in a more intuitive manner such as using the multiparameter squeezing parameter \cite{gessner2019} other than entanglement, which we leave for future study.
Finally, finding the optimal measurement achieving the best sensitivity in the presence of photon-loss is also an important remaining task as in the recent study, where the optimal non-Gaussian measurement in a special case of $M=1$ is theoretically found \cite{oh2019-1}. The experimental implementation of the optimal measurement needs to be devised.

It is worthwhile to discuss our results in relation to recent results in distributed sensing.
First of all, a recent experiment successfully showed an enhancement by entanglement in distributed Gaussian quantum sensing~\cite{guo2019}. The theory behind the experiment in Ref.~\cite{guo2019} assumed that the phase shifts of interest were extremely small and the estimation error was quantified by the linear error propagation analysis from homodyne detection. 
However, our work identifies the ultimate estimation error in distributed Gaussian sensing by proposing the optimal Gaussian probe and it can be applied to phase shifts of arbitrary degrees. 
Thus, the experimental results could be understood better and interpreted from a broader perspective of distributed Gaussian sensing.
In addition, the optimal entangled Gaussian state has been proven to be optimal for distributed quantum sensing of field-quadrature displacement \cite{zhuang2018, xia2019, zhuang2019}.

%\paragraph{Acknowledgements.---}
This work was supported by National Research Foundation of Korea (NRF) grants funded by the Korea government  (Grants No. NRF-2019R1H1A3079890 and No. NRF-2018K2A9A1A06069933).

C.O. and C.L. contributed equally to this work.

\onecolumngrid

\setcounter{equation}{0}
\renewcommand{\theequation}{A\arabic{equation}}

\appendix
\section{Derivation of the quantum Fisher information matrix for distributed sensing using isothermal Gaussian states} \label{appA}
In this section, we derive the quantum Fisher information matrix (QFIM) for distributed sensing using isothermal Gaussian states.
When a phase-encoded state is the isothermal Gaussian quantum states characterized by $[\bm{\Gamma}(\bm{\phi}),\bm{d}(\bm{\phi})]$ with a isothermal photon number $\bar{n}$, the QFIM is given by \cite{oh2019-2}
\begin{align}\label{eq:H_ij}
H_{ij}=\frac{1}{2\bar{n}^2+2\bar{n}+1}\text{Tr}\left[\bm{\Omega}_{2M}\frac{\partial \bm{\Gamma}(\bm{\phi})}{\partial\phi_i}\bm{\Omega}_{2M}\frac{\partial \bm{\Gamma}(\bm{\phi})}{\partial\phi_j}\right]+\frac{\partial \bm{d}^\text{T}(\bm{\phi})}{\partial \phi_i}\bm{\Gamma}^{-1}\frac{\partial \bm{d}(\bm{\phi})}{\partial \phi_j},
\end{align}
where 
\begin{align}
\bm{\Gamma}(\bm{\phi})=S(\bm{\phi})\bm{\Gamma}_\text{probe} S^\text{T}(\bm{\phi}), \quad
\bm{d}(\bm{\phi})=S(\bm{\phi})\bm{d}_\text{probe},\nonumber
\end{align}
are the covariance matrix and the first moment vector of the quantum state after the unitary operation encoding $\bm{\phi}$ corresponding to the symplectic matrix $S(\bm{\phi})$, respectively.
In the distributed phase sensor, the symplectic transformation corresponds to
\begin{align}
S(\bm{\phi})=\oplus_{i=1}^M
\begin{pmatrix}
\cos \phi_i & \sin \phi_i \\
-\sin \phi_i & \cos \phi_i
\end{pmatrix}.\nonumber
\end{align}
Note that symplectic transformation $S$ is defined as ones that preserve the canonical commutation relation, $S^\text{T}\bm{\Omega}_{2M} S=\bm{\Omega}_{2M}$, corresponding to a Gaussian unitary operation $\hat{U}$ applied to density matrices by the relation $\hat{U}^\dagger\hat{Q}\hat{U}=S\hat{Q}$.

The first term in Eq.~\eqref{eq:H_ij} can be simplified as
\begin{align}
\text{Tr}\left[\bm{\Omega}_{2M}\frac{\partial \bm{\Gamma}}{\partial\phi_i}\bm{\Omega}_{2M}\frac{\partial \bm{\Gamma}}{\partial\phi_j}\right]
=&\text{Tr}\bigg[\bm{\Omega}_{2M}\frac{\partial S(\bm{\phi})}{\partial \phi_i}\bm{\Gamma}_\text{probe}\bm{\Omega}_{2M}\frac{\partial S(\bm{\phi})}{\partial \phi_j}\bm{\Gamma}_\text{probe}
+\bm{\Omega}_{2M}\frac{\partial S(\bm{\phi})}{\partial \phi_i}\bm{\Gamma}_\text{probe}\bm{\Omega}_{2M}\bm{\Gamma}_\text{probe}\frac{\partial S^\text{T}(\bm{\phi})}{\partial \phi_j} \nonumber \\
&\quad+\bm{\Omega}_{2M}\bm{\Gamma}_\text{probe}\frac{\partial S^\text{T}(\bm{\phi})}{\partial \phi_i}\bm{\Omega}_{2M}\frac{\partial S(\bm{\phi})}{\partial \phi_j}\bm{\Gamma}_\text{probe}
+\bm{\Omega}_{2M}\bm{\Gamma}_\text{probe}\frac{\partial S^\text{T}(\bm{\phi})}{\partial \phi_i}\bm{\Omega}_{2M}\bm{\Gamma}_\text{probe}\frac{\partial S^\text{T}(\bm{\phi})}{\partial \phi_j}\bigg] \nonumber\\
\overset{\bm{\phi}=0}{=}&\text{Tr}[P_i\bm{\Gamma}_\text{probe}P_j\bm{\Gamma}_\text{probe}+P_i\bm{\Gamma}_\text{probe}\bm{\Omega}_{2M}\bm{\Gamma}_\text{probe}\bm{\Omega}_{2M} P_j+P_j\bm{\Gamma}_\text{probe}\bm{\Omega}_{2M}\bm{\Gamma}_\text{probe}\bm{\Omega}_{2M} P_i \nonumber \\
&\quad+\bm{\Gamma}_\text{probe}P_i\bm{\Gamma}_\text{probe}P_j] \nonumber\\
=&2\text{Tr}[\bm{\Gamma}^{(i,j)}_\text{probe}\bm{\Gamma}^{(j,i)}_\text{probe}]-\delta_{ij}(2\bar{n}+1)^2\label{eq:1st_term}
\end{align}
where $\bm{\Gamma}^{(i,j)}_\text{probe}=P_i\bm{\Gamma}_\text{probe}P_j$. Here, we have set $\bm{\phi}=0$ without loss of generality since the QFIM is independent of $\bm{\phi}$ under unitary transformation, and we have used
\begin{align}
\frac{\partial \bm{\Gamma}(\bm{\phi})}{\partial \phi_i}&=\frac{\partial S(\bm{\phi})}{\partial \phi_i}\bm{\Gamma}_\text{probe} S^\text{T}(\bm{\phi})+S(\bm{\phi})\bm{\Gamma}_\text{probe} \frac{\partial S^\text{T}(\bm{\phi})}{\partial \phi_i}=\frac{\partial S(\bm{\phi})}{\partial \phi_i}\bm{\Gamma}_\text{probe}+\bm{\Gamma}_\text{probe} \frac{\partial S^\text{T}(\bm{\phi})}{\partial \phi_i}, \nonumber
\end{align}
and
\begin{align}
-\bm{\Omega}_{2M}\frac{\partial S(\bm{\phi})}{\partial\phi_i}&=-\frac{\partial S(\bm{\phi})}{\partial\phi_i}\bm{\Omega}_{2M}=\bm{\Omega}_{2M}\frac{\partial S^\text{T}(\bm{\phi})}{\partial\phi_i}=\frac{\partial S^\text{T}(\bm{\phi})}{\partial\phi_i}\bm{\Omega}_{2M}, \nonumber
%\label{projection}
\end{align}
which is the projection onto the $i$th mode, $P_i=-\bm{\Omega}_{2M}\frac{\partial S(\bm{\phi})}{\partial\phi_i}\bigg|_{\bm{\phi}=0}$ when $\bm{\phi}=0$.

The second term in Eq.~\eqref{eq:H_ij} is
\begin{align}
\frac{\partial \bm{d}^\text{T}(\bm{\phi})}{\partial \phi_i}\bm{\Gamma}^{-1}\frac{\partial \bm{d}(\bm{\phi})}{\partial \phi_j}=\left(\frac{\partial S(\bm{\phi})}{\partial \phi_i}\bm{d}_\text{probe}\right)^\text{T}\bm{\Gamma}^{-1}\left(\frac{\partial S(\bm{\phi})}{\partial \phi_j}\bm{d}_\text{probe}\right)\overset{\bm{\phi}=0}{=}
(\bm{\Omega}_2 \bm{d}^{(i)}_\text{probe})^\text{T}[\bm{\Gamma}^{-1}]^{(i,j)}(\bm{\Omega}_2 \bm{d}^{(i)}_\text{probe})\label{eq:2nd_term}.
\end{align}
Thus, substituting $\bar{n}=0$ into Eqs.~\eqref{eq:H_ij}$\sim$\eqref{eq:2nd_term}, i.e., for pure states and using $\bm{\Gamma}_\text{probe}=\bm{\Gamma}$ and $\bm{d}_\text{probe}=\bm{d}$ since we have set $\bm{\phi}=0$, we obtain the expression of Eq.~(2) in the main text. 

\setcounter{equation}{0}
\renewcommand{\theequation}{B\arabic{equation}}

\section{Maximum variance of $\hat{G}^*$} \label{appB}
Let us derive the maximum variance of $\hat{G}^*=\sum_{i=1}^M \hat{a}_i^\dagger\hat{a}_i/M$. The variance can be written as
\begin{align}
M^2(\Delta^2 \hat{G}^*)=M^2(\langle\hat{G}^{*2}\rangle-\langle\hat{G}^*\rangle^2)
&=\langle \left(\sum_{i=1}^M \hat{a}_i^\dagger\hat{a}_i\right)^2\rangle-\left(\sum_{i=1}^M \langle \hat{a}_i^\dagger\hat{a}_i\rangle\right)^2\nonumber\\
&= \sum_{i=1}^M\langle(\hat{a}_i^\dagger\hat{a}_i)^2\rangle+\sum_{i\neq j}^M\langle\hat{a}_i^\dagger\hat{a}_i\hat{a}_j^\dagger\hat{a}_j\rangle-\sum_{i=1}^M \langle \hat{a}_i^\dagger\hat{a}_i\rangle^2-\sum_{i\neq j}^M \langle \hat{a}_i^\dagger\hat{a}_i\rangle\langle \hat{a}_j^\dagger\hat{a}_j\rangle.\nonumber
\end{align}
Using the fact that $\hat{G}^*$ is invariant under any passive transformation, one can assume that $\langle\hat{a}_i^\dagger\hat{a}_i\hat{a}_j^\dagger\hat{a}_j\rangle-\langle \hat{a}_i^\dagger\hat{a}_i\rangle\langle \hat{a}_j^\dagger\hat{a}_j\rangle=0$ for $i\neq j$ without loss of generality and get 
\begin{align}
\Delta^2 \hat{G}^*=\frac{1}{M^2}\sum_{i=1}^M \Delta^2 (\hat{a}_i^\dagger\hat{a}_i)\nonumber,
\end{align}
which shows the variance of $\Delta^2 \hat{G}^*$ is the sum of the photon number variance in all the modes. Since a squeezed vacuum state exhibits the maximum photon number variance among Gaussian states, which is $(\cosh 4r-1)/4$, 
\begin{align}
\Delta^2 \hat{G}^*\leq \frac{1}{4M^2}\sum_{i=1}^M (\cosh 4r_i-1).\label{varG_max}
\end{align}
Under the constraint for the total mean photon number of the state $\bar{N}$, one can prove that the upper bound of $\Delta^2 \hat{G}^*$ in Eq.~\eqref{varG_max} is given by $2\bar{N}(\bar{N}+1)$, i.e.,
\begin{align}
4\Delta^2 \hat{G}^*\leq \frac{8\bar{N}(\bar{N}+1)}{M^2}.
\end{align}

\setcounter{equation}{0}
\renewcommand{\theequation}{C\arabic{equation}}

\section{Properties of the QFIM for symmetric Gaussian states} \label{appC}
Let us consider the QFIM having diagonal elements $H_{11}$ and off-diagonal elements $H_{12}$, which is then written as
\begin{align}
\bm{H}
&=H_{11}\sum_{i=1}^M|i\rangle\langle i|+H_{12}\sum_{i\neq j}^M |i\rangle\langle j|\nonumber\\
&=(H_{11}-H_{12})\sum_{i=1}^M|i\rangle\langle i|+H_{12}\sum_{i, j=1}^M |i\rangle\langle j|\nonumber\\
&=(H_{11}-H_{12})\mathbb{1}+H_{12}\sum_{i, j=1}^M |i\rangle\langle j|\nonumber,
\end{align}
where $\{|i\rangle\}_{i=1}^M$ represents the standard basis.
By introducing $|+\rangle=\sum_{i=1}^M|i\rangle/\sqrt{M}$,
\begin{align}
\bm{H}&=(H_{11}-H_{12})\mathbb{1}+MH_{12}|+\rangle\langle +|\nonumber\\
&=(H_{11}-H_{12})(\mathbb{1}-|+\rangle\langle+|)+[(M-1)H_{12}+H_{11}]|+\rangle\langle +|.\nonumber
\end{align}
When $H_{11}\neq H_{12}$, the inverse of the QFIM is
\begin{align}
\bm{H}^{-1}=(H_{11}-H_{12})^{-1}(\mathbb{1}-|+\rangle\langle+|)+[(M-1)H_{12}+H_{11}]^{-1}|+\rangle\langle +|,\nonumber
\end{align}
and thus 
\begin{align}
\text{Tr}[\bm{H}^{-1}]=(M-1)(H_{11}-H_{12})^{-1}+[(M-1)H_{12}+H_{11}]^{-1}.\nonumber
\end{align}
When $H_{11}=H_{12}$, however, the inverse of the QFIM is
\begin{align}
\bm{H}^{-1}=[(M-1)H_{12}+H_{11}]^{-1}|+\rangle\langle +|,\nonumber
\end{align}
and thus
\begin{align}
\text{Tr}[\bm{H}^{-1}]=[(M-1)H_{12}+H_{11}]^{-1}.\nonumber
\end{align}

\setcounter{equation}{0}
\renewcommand{\theequation}{D\arabic{equation}}

\section{Minimization of the estimation error when probing with symmetric Gaussian states} \label{appD}
For pure symmetric Gaussian states, the elements of the covariance matrix satisfy
\begin{align}
(\gamma_1-\epsilon_1)(\gamma_2-\epsilon_2)=1/4, \nonumber\\
[\gamma_1+(M-1)\epsilon_1][\gamma_2+(M-1)\epsilon_2]=1/4,\nonumber
\end{align}
and the energy constraint is
\begin{align}
\bar{N}=M(\gamma_1+\gamma_2-1)/2.\nonumber
\end{align}
Parametrizing $\gamma_{1,2}$ as
\begin{align}
\gamma_{1,2}=\bar{n}_\text{T} e^{\pm 2r},\nonumber
\end{align}
we can rewrite $\epsilon_{1,2}$ as
\begin{align}
\epsilon_{1,2}=\frac{2+4\bar{n}_\text{T}^2(M-2)-M\pm\sqrt{(4\bar{n}_\text{T}^2-1)[M(4\bar{n}_\text{T}^2 M-M+4)-4]}}{8\bar{n}_\text{T}(M-1)}e^{\pm2r},\nonumber
\end{align}
where $0\leq r \leq \cosh^{-1}(2\bar{N}/M+1)/2$.

Under the above constraint, our task boils down to finding parameters that maximize
\begin{align}
M[2(\gamma_1^2+\gamma_2^2)-1+2(M-1)(\epsilon_1^2+\epsilon_2^2)],\label{eq:term_TBM}
\end{align}
whose maximum value can be shown to be $8\bar{N}(\bar{N}+1)$ in general.

One can easily check that if we use
\begin{align}
\gamma_{1,2}=\frac{1}{2}+\frac{\bar{N}\pm\sqrt{\bar{N}(\bar{N}+1)}}{M},\label{eq:gamma}
\end{align}
and
\begin{align}
\epsilon_{1,2}=\frac{\bar{N}\pm\sqrt{\bar{N}(\bar{N}+1)}}{M},\label{eq:epsilon}
\end{align}
the maximum value of Eq.~\eqref{eq:term_TBM}, i.e., $8\bar{N}(\bar{N}+1)$, is attained, which thus proves that $\Delta^2 \phi^*_\text{OEGS}$ introduced in the main text is achievable by the symmetric Gaussian states with parameters satisfying Eqs.~\eqref{eq:gamma} and \eqref{eq:epsilon}.

\setcounter{equation}{0}
\renewcommand{\theequation}{E\arabic{equation}}

\section{General Gaussian measurement}
In this section, we derive the lower bound of the estimation error based on a particular Gaussian measurement and provide its implementation.
A measurement is called a Gaussian measurement if it can be implemented by adding Gaussian ancilla states with Gaussian unitary operations and performing homodyne detection \cite{weedbrook2012, serafini2017}.
Mathematically, a Gaussian measurement on $M$-mode states $\hat{\rho}$ can be written by positive-valued measure measure (POVM) elements $\{\hat{\Pi}_{\bm{\xi}}\}$ as
\begin{align}
\hat{\Pi}_{\bm{\xi}}=\frac{1}{\pi^M}\hat{D}(\bm{\xi})\hat{\Pi}_0\hat{D}^\dagger(\bm{\xi}),\nonumber
\end{align}
where $\hat{D}(\bm{\xi})=\exp(-i\bm{\xi}^\text{T}\Omega_{2M} \hat{\bm{Q}})$ is a displacement operator, and $\hat{\Pi}_0$ is a density matrix of a $M$-mode Gaussian state with a zero-displacement and a covariance matrix $\bm{\Gamma}_\text{M}$. Note here that $\hat{\Pi}_0$ characterizes the Gaussian measurement. Let us assume $\hat{\Pi}_0$ to be a pure state.
%the output probability distribution for $\bm{\xi}$ is given by $p(\bm{\xi})=\text{Tr}[\hat{\rho}\hat{\Pi}_{\bm{\xi}}]$.
One can easily show that the probability distribution for a Gaussian input state with the covariance matrix $\bm{\Gamma}$ and the first moment $\bm{d}$ is given as a Gaussian distribution with the covariance matrix $(\bm{\Gamma}+\bm{\Gamma}_\text{M})/2$ and the first moment $\bm{d}/\sqrt{2}$.
For the phase-encoded Gaussian state of $\Gamma(\bm{\phi})$ with zero displacement, the Fisher information elements based on Gaussian measurement with $\bm{\Gamma}_\text{M}$ are thus given by
\begin{align}
F_{ij}(\bm{\phi})=\frac{1}{2}\text{Tr}\left[(\bm{\Gamma}+\bm{\Gamma}_\text{M})^{-1}\frac{\partial\bm{\Gamma}}{\partial\phi_i}(\bm{\Gamma}+\bm{\Gamma}_\text{M})^{-1}\frac{\partial\Gamma}{\partial\phi_j}\right].\nonumber
\end{align}

Let us consider a Gaussian measurement $\hat{\Pi}_0$ with the following covariance matrix:
\begin{align}
\bm{\Gamma}_\text{M}=
\begin{pmatrix}
\bm{\gamma}_\text{M} & \bm{\epsilon}_\text{M} & ... & \bm{\epsilon}_\text{M} \\ 
\bm{\epsilon}_\text{M} & \bm{\gamma}_\text{M} & ... & \bm{\epsilon}_\text{M} \\ 
\vdots & \vdots & ... & \vdots \\
\bm{\epsilon}_\text{M} & \bm{\epsilon}_\text{M} & ... & \bm{\gamma}_\text{M}
\end{pmatrix},\nonumber
\end{align}
where $\bm{\gamma_\text{M}}=\text{diag}(\gamma_{\text{M},1},\gamma_{\text{M},2})$ and $\bm{\epsilon}_\text{M}=\text{diag}(\epsilon_{\text{M},1},\epsilon_{\text{M},2})$ are $2\times 2$ diagonal matrices, and
$\gamma_{\text{M},j}=1/2+\epsilon_{\text{M},j}$ and
$\epsilon_{\text{M},j}=[\bar{N}_\text{M}-(-1)^j\sqrt{\bar{N}_\text{M}(\bar{N}_\text{M}+1)}]/M$ for $j=1, 2$.
Note that the covariance matrix is the same as that of the optimal entangled Gaussian state with $\bar{N}$ replaced by $\bar{N}_\text{M}$.
If the phase-encoded state is the optimal entangled Gaussian state in the presence of loss, then one can find that the lower bound of the error can be written as
\begin{align}
\Delta^2 \phi^*&\geq\frac{2 \eta  \bar{N}_\text{M} \bar{N}-2 \eta  \sqrt{\bar{N}_\text{M} (\bar{N}_\text{M}+1) \bar{N} (\bar{N}+1)}+\bar{N}_\text{M}-(\eta -2) \eta  \bar{N}+1}{4 \eta ^2 \bar{N} (\bar{N}+1)} \nonumber\\
&\geq \frac{2\bar{N} (1-\eta) \eta+1+\sqrt{1-4\bar{N} \eta(\eta -1)}}{8 \eta ^2 \bar{N} (\bar{N}+1)},\nonumber
\end{align}
where the optimal value of $\bar{N}_\text{M}$ is chosen for the second inequality.

If we employ a squeezed thermal input state,~$\hat{\rho}_\text{in}=\hat{S}(r) \hat{\rho}_\text{T} \hat{S}^\dagger(r)\otimes |0\rangle\langle0|^{\otimes M-1}$ where~$\hat{S}(r)=\exp[r(\hat{a}^{\dagger2}-\hat{a}^2)/2]$ is a squeezing operator applied on the first mode, and~$\hat{\rho}_\text{T}=\sum_{n=0}^\infty \bar{n}^n/(\bar{n}+1)^{n+1}|n\rangle\langle n|$ is a thermal state with the mean photon number~$\bar{n}$, the lower bound by the aforementioned Gaussian measurement can be written as
\begin{align}
\Delta^2 \phi^*\geq \left[\left(\frac{2\bar{n}+1}{\bar{n}+1}\right)^2\sinh^2{2r}\right]^{-1},\nonumber
\end{align}
which is exactly the same as the lower bound for single-mode phase estimation using a squeezed thermal probe state, as shown in Ref.~\cite{oh2019-1}.
Note that preparing the squeezed thermal state input without a photon-loss channel is equivalent to using the optimal entangled Gaussian state with a photon-loss channel after adjusting appropriate parameters when the photon-loss rates are equal to each other \cite{oh2017}.

Let us find the implementation of the Gaussian measurement corresponding to $\Gamma_\text{M}$.
Noticing that mixing a $p$-squeezed state and $(M-1)$ vacua by the first beam splitter network (BSN) in the main text generates the optimal entangled state, $\hat{\Pi}_0$ can be represented by
\begin{align}
\hat{\Pi}_0=\hat{U}_\text{BSN} |r,0,...,0\rangle \langle r,0,...,0|\hat{U}_\text{BSN}^\dagger,\nonumber
\end{align}
where $|r\rangle\langle r|$ represents a $p$-squeezed state with a squeezing parameter $r$ and $|0\rangle\langle 0|$ is a vacuum state.
Since a BSN transforms a displacement operator into another displacement operator and a single-mode Gaussian measurement of $\hat{\Pi}_{\zeta}=\hat{D}(\bm{\zeta})\hat{\Pi}_0\hat{D}^\dagger(\bm{\zeta})/\pi$ can be implemented by the general-dyne measurement \cite{genoni2014},
the Gaussian measurement can be performed by general-dyne measurement on $M$ modes after the second BSN that processes the reverse of the first BSN generating the optimal entangled state.
Especially when $\hat{\Pi}_0$ is a vacuum, the general-dyne measurement reduces to a heterodyne measurement.

\twocolumngrid


\begin{references}
%Quantum metrology
%\bibitem{caves1981} C. M. Caves, Quantum-mechanical noise in an interferometer, Phys. Rev. D {\bf 23}, 1693 (1981).
%\bibitem{bollinger1996} J. J. Bollinger, W. M. Itano, D. J. Wineland, and D. J. Heinzen, Optimal frequency measurements with maximally correlated states, Phys. Rev. A {\bf 54}, R4649 (1996).
\bibitem{giovannetti2004} V. Giovannetti, S. Lloyd, and L. Maccone, Quantum-Enhanced Measurements: Beating the Standard Quantum Limit, Science {\bf 306}, 1330 (2004).
\bibitem{giovannetti2011} V. Giovannetti, S. Lloyd, and L. Maccone, Advances in quantum metrology, Nat. Photon. {\bf 5,} 222–229 (2011).
\bibitem{demko2015} R. Demkowicz-Dobrzanski, M. Jarzyna, J. Kolodynski, Quantum limits in optical interferometry, Progress in Optics {\bf 60}, 345–435 (2015).

\bibitem{pirandola2018} S. Pirandola, B. R. Bardhan, T. Gehring, C. Weedbrook, and S. Lloyd, Advances in photonic quantum sensing
, Nat. Photon. {\bf 12,} 724-733 (2018).
\bibitem{braun2018} D. Braun, G. Adesso, F. Benatti, R. Floreanini, U. Marzolino, M. W. Mitchell, and S. Pirandola, Quantum-enhanced measurements without entanglement, Rev. Mod. Phys. {\bf 90,} 035006, (2018).



\bibitem{Degen2017} C. L. Degen, F. Reinhard, and P. Cappellaro, Quantum sensing, Rev. Mod. Phys. {\bf 89}, 035002 (2017).

\bibitem{Szczykulska2016} M. Szczykulska, T. Baumgratz, and A. Datta, Multi-parameter quantum metrology, Adv. Phys. X: {\bf 1}, 621 (2016).

%Multiparameter estimation
\bibitem{humphreys2013} P. C. Humphreys, M. Barbieri, A. Datta, and I. A. Walmsley, Quantum Enhanced Multiple Phase Estimation, \prl~{\bf 111}, 070403 (2013).
\bibitem{Liberman2015} L. Liberman, Y. Israel, E. Poem, and Y. Silberberg, Quantum enhanced phase retrieval, Optica {\bf 3}, 193 (2016).
\bibitem{baumgratz2016} T. Baumgratz and A. Datta, Quantum Enhanced Estimation of a Multidimensional Field, Phys. Rev. Lett. {\bf 116}, 030801 (2016).
\bibitem{Knott2016} P. A. Knott, T. J. Proctor, A. J. Hayes, J. F. Ralph, P. Kok, and J. A. Dunningham, Local versus global strategies in multiparameter estimation, Phys. Rev. A {\bf 94}, 062312 (2016).


%distributed sensor
\bibitem{pezze2017} L. Pezze, M. A. Ciampini, N. Spagnolo, P. C. Humphreys, A. Datta, I. A. Walmsley, M. Barbieri, F. Sciarrino, and A. Smerzi, Optimal Measurements for Simultaneous Quantum Estimation of Multiple Phases, \prl~ {\bf 119}, 130504 (2017).
\bibitem{proctor2018} T. J. Proctor, P. A. Knott, and J. A. Dunningham, Multiparameter Estimation in Networked Quantum Sensors, \prl~ {\bf 120}, 080501 (2018).
\bibitem{ge2018} W. Ge, K. Jacobs, Z. Eldredge, A. V. Gorshkov and M. Foss-Feig, Distributed Quantum Metrology with Linear Networks and Separable Inputs, \prl~ {\bf 121}, 043604 (2018).
\bibitem{gessner2018} M. Gessner, L. Pezz{\`e}, and A. Smerzi, Sensitivity Bounds for Multiparameter Quantum Metrology, Phys. Rev. Lett. {\bf 121}, 130503 (2018).
\bibitem{guo2019} X. Guo, C. R. Breum, J. Borregaard, S. Izumi, M. V. Larsen, M. Christandl, J. S. Neergaard-Nielsen, and U. L. Andersen, Distributed quantum sensing in a continuous-variable entangled network, arXiv:1905.09408 (2019).
\bibitem{gatto2019} D. Gatto, P. Facchi, F. Narducci, and V. Tamma, Distributed quantum metrology with a single squeezed-vacuum source, Phys. Rev. Research {\bf 1,} 032024 (R) (2019).

\bibitem{gessner2019} M. Gessner, A. Smerzi, and L. Pezze, Metrological Multiparameter Squeezing, arXiv:1910.14014

%global sensing
\bibitem{komar2014} P. Komar, E.M. Kessler, M. Bishof, L. Jiang, A. S. S{\o}rensen, J. Ye, and M. D. Lukin, A quantum network of clocks, Nat. Phys. {\bf 10}, 582 (2014).

\bibitem{Gagatsos2016} C. N. Gagatsos, D. Branford, and A. Datta, Gaussian systems for quantum-enhanced multiple phase estimation, Phys. Rev. A {\bf 94}, 042342 (2016).


%Gaussian states
\bibitem{ferraro2005} A. Ferraro, S. Olivares, and M. G. A. Paris, Gaussian States in Quantum Information (Bibliopolis, Berkeley, 2005).
%\bibitem{wang2007} X.-B. Wang, T. Hiroshima, A. Tomita, and M. Hayashi, Quantum information with Gaussian states, Phys. Rep. {\bf 448}, 1 (2007).
\bibitem{weedbrook2012} C. Weedbrook, S. Pirandola, R. Garcia-Patron, N. J. Cerf, T. C. Ralph, J. H. Shapiro, and S. Lloyd, Gaussian quantum information, Rev. Mod. Phys. {\bf 84}, 621 (2012).
\bibitem{adesso2014} G. Adesso, S. Ragy, and A. R. Lee, Continuous variable quantum information: Gaussian states and beyond, Open Syst. Inf. Dyn. {\bf 21}, 1440001 (2014).


%QFIM
\bibitem{helstrom1976} C. W. Helstrom, Mathematics in Science and Engineering (Academic Press, New York, 1976), Vol. 123.
\bibitem{braunstein1994} S. L. Braunstein and C. M. Caves, Statistical Distance and the Geometry of Quantum States, Phys. Rev. Lett. {\bf 72}, 3439 (1994).
\bibitem{paris2009} M. G. A. Paris, Quantum estimation for quantum technology, Int. J. Quantum Inf. {\bf 7}, 125 (2009).

\bibitem{Reck1994} M. Reck, A. Zeilinger, H. J. Bernstein, and P. Bertani, Experimental realization of any discrete unitary operator, Phys. Rev. Lett. {\bf 73}, 58 (1994).
\bibitem{jarzyna2012} M. Jarzyna and R. Demkowicz-Dobrza{\' n}ski, Quantum interferometry with and without an external phase reference, Phys. Rev. A {\bf 85}, 011801(R) (2012).


\bibitem{banchi2015} L. Banchi, S. L. Braunstein, and S. Pirandola, Quantum Fidelity for Arbitrary Gaussian States, \prl~ {\bf 115,} 260501 (2015).

\bibitem{serafini2017} A. Serafini, {\it Quantum Continuous Variables: A Primer of Theoretical Methods} (Taylor \& Francis, Oxford, 2017).

\bibitem{nichols2018}  R. Nichols, P. Liuzzo-Scorpo, P. A. Knott, and G. Adesso, Multiparameter Gaussian quantum metrology, Phys. Rev. A {\bf 98}, 012114 (2018)
\bibitem{oh2019-2} C. Oh, C. Lee, L. Banchi, S.-Y. Lee, C. Rockstuhl, and H. Jeong, Optimal measurements for quantum fidelity between Gaussian states and its relevance to quantum metrology, Phys. Rev. A {\bf 100}, 012323 (2019).
\bibitem{sidhu2019} J. S. Sidhu and P. Kok, Geometric Perspective on Quantum Parameter Estimation,  AVS Quantum Science {\bf 2,} 014701 (2020).
\bibitem{liu2019} J. Liu, H. Yuan, X.-M. Lu, X. Wang, Quantum Fisher information matrix and multiparameter estimation, J. Phys. A: Math. Theor. 53, 023001 (2020).

\bibitem{olivares2009} S. Olivares and M. G. A. Paris, Bayesian estimation in homodyne interferometry, J. Phys. B. {\bf 42}, 055506 (2009). 



%symmetric gaussian state
\bibitem{adesso2004} G. Adesso, A. Serafini, and F. Illuminati, Quantification and Scaling of Multipartite Entanglement in Continuous Variable Systems, \prl~ {\bf 93}, 220504 (2004).
\bibitem{serafini2005} A. Serafini, G. Adesso, and F. Illuminati, Unitarily localizable entanglement of Gaussian states, \pra~ {\bf 71}, 032349 (2005).
\bibitem{adesso2008} G. Adesso and F. Illuminati, Genuine multipartite entanglement of symmetric Gaussian states: Strong monogamy, unitary localization, scaling behavior, and molecular sharing structure, \pra~ {\bf 78}, 042310 (2008).

\bibitem{adesso2005} G. Adesso and F. Illuminati, Equivalence between Entanglement and the Optimal Fidelity of Continuous Variable Teleportation, \prl~ {\bf 95}, 150503 (2005).

%State generation
\bibitem{loock2000} P. van Loock and S. L. Braunstein, Multipartite Entanglement for Continuous Variables: A Quantum Teleportation Network, \prl~{\bf 84}, 3482 (2000).
\bibitem{loock2003} P. van Loock and A. Furusawa, Detecting genuine multipartite continuous-variable entanglement, \pra~{\bf 67}, 052315 (2003).
\bibitem{yonezawa2004} H. Yonezawa, A. Takao, and A. Furusawa. Demonstration of a quantum teleportation network for continuous variables, Nature {\bf 431}, 430 (2004).



%\bibitem{modi2010} K. Modi, T. Paterek, W. Son, V. Vedral, and M. Williamson, Unified View of Quantum and Classical Correlations, \prl~ {\bf 104}, 080501 (2010).


%non-Gaussian
\bibitem{oh2019-1} C. Oh, C. Lee, C. Rockstuhl, H. Jeong, J. Kim, H. Nha, S.-Y. Lee, Optimal Gaussian measurements for phase estimation in single-mode Gaussian metrology, npj Quantum Inf. {\bf 5}, 10 (2019).

%distributed displacement sensing

\bibitem{zhuang2018} Q. Zhuang, Z. Zhang, and J. H. Shapiro, Distributed quantum sensing using continuous-variable multipartite entanglement, \pra~ {\bf 97,} 032329 (2018).
\bibitem{xia2019} Y. Xia, Q. Zhuang, W. Clark, and Z. Zhang, Repeater-enhanced distributed quantum sensing based on continuous-variable multipartite entanglement, \pra~ {\bf 99,} 012328 (2019).
\bibitem{zhuang2019} Q. Zhuang, J. Preskill, and L. Jiang, Distributed quantum sensing enhanced by continuous-variable error correction, arXiv:1910.14156 (2019).
\bibitem{oh2017} C. Oh, S.-Y. Lee, H. Nha, and H. Jeong, Practical resources and measurements for lossy optical quantum metrology, Phys. Rev. A {\bf 96,} 062304 (2017).



\bibitem{genoni2014} M. G. Genoni, S. Mancini, and A. Serafini, General-dyne unravelling of a thermal master equation. Russ. J. Math. Phys. {\bf 21,} 329 (2014).


%appendix
%\bibitem{genoni2014} M. G. Genoni, S. Mancini, and A. Serafini, General-dyne unravelling of a thermal master equation. Russ. J. Math. Phys. {\bf 21}, 329 (2014).

\end{references}
\end{document}